# Observation of enhanced transmission for s-polarized light through a subwavelength slit

M. Guillaumée,[1,a)] A. Yu. Nikitin,[2,3] M. J. K. Klein,[1] L. A. Dunbar,[1] V. Spassov,[1] R. Eckert,[1] L. Martín-Moreno,[2], F. J. García-Vidal[4] and R. P. Stanley[1]

[1]*Swiss Centre for Electronics and Microtechnology, CSEM SA, Jaquet-Droz 1, CH-2002 Neuchâtel, Switzerland*

[2]*Instituto de Ciencia de Materiales de Aragon and Departamento de Fisica de la Materia Condensada, CSIC-Universidad de Zaragoza, E-50009, Zaragoza, Spain*

[3]*A. Ya. Usikov Institute for Radiophysics and Electronics, Ukrainian Academy of Sciences, 61085 Kharkov, Ukraine*

[4]*Departamento de Fisica Teorica de la Materia Condensada,Universidad Autonoma de Madrid, E-28049 Madrid, Spain*

Enhanced optical transmission (EOT) through subwavelength apertures is usually obtained for p-polarized light. The present study experimentally investigates EOT for s-polarized light. A subwavelength slit surrounded on each side by periodic grooves has been fabricated in a gold film and covered by a thin dielectric layer. The excitation of s-polarized dielectric waveguide modes inside the dielectric film strongly increases the s-polarized transmission. Transmission measurements are compared with a coupled mode model and show good qualitative agreement. Adding a waveguide can improve light transmission through subwavelength apertures, as both s and p-polarization can be efficiently transmitted.

---

a) Electronic mail: mickael.guillaumee@csem.ch.

Light transmission through single subwavelength apertures drilled in a metallic film can be greatly increased by periodically structuring the surface surrounding the aperture.[1,2] The normalized transmission $\eta$, defined as the transmission through the structure normalized to the photon flux incident on an open aperture, can be several times larger than unity. This high transmission phenomenon, called extraordinary optical transmission, has been achieved with the so called bull's eye structure, or its one dimensional equivalent, a single slit flanked by periodic grooves on each side.[2] The grooves resonantly couple incident light into surface waves, which constructively interfere with the light directly incident on the aperture.[3,4] The surface waves are either surface plasmon polaritons (SPP)[5] in the case of real metals in the visible and near infrared spectral range, or spoof SPP for perfect metals.[6] Both conventional or spoof SPPs are p-polarized waves, i.e. the magnetic field is parallel to the grooves. In the slit and groove case, the absence of cutoff frequency of the fundamental p-polarized slit mode renders possible the excitation of cavity resonances for any slit width,[7] increasing further light transmission.[3] As both waveguide resonances in subwavelength slits and excitation of surface waves are p-polarized processes, slit and grooves structures are intrinsically polarization sensitive. This has been advantageously exploited in several studies, for example to filter polarization[8] or to control the polarization state of optical devices.[9] However, most of the s-polarized light is lost. This is a drawback in applications where high throughput is necessary, such as low noise[8] or high speed[10] photodetectors.

In order to increase s-polarized transmission, it is first necessary to use a slit width such that it is above the cutoff width. Also, the presence of a thin dielectric layer on top of the slit and groove structure permits s-polarized incident light to couple to dielectric waveguide modes, be directed toward the slit and boost s-polarized transmission. At the same time, p-polarized light is still resonantly transmitted. This scheme was recently proposed and theoretically investigated by several of the present authors.[11]

The goal of the present paper is to validate this concept experimentally. A dielectric layer is fabricated on top of a slit and groove structure. The experimental measurements show that a dielectric layer radically increases s-polarized transmission. Comparisons with theoretical calculations are reported.

Each structure is composed of a single slit with 7 periodic grooves on either side. The grooves are 180 nm deep and 325 nm wide. Both the slits and grooves are 10 μm long. The slit width $w$ and the period $p$ are the parameters varied in this study. A 250 nm thin dielectric layer covers the slit and groove structure. A schematic view of a slit and groove structure covered with a thin dielectric film is shown in Fig. 1(a).

First, a 420 ± 10 nm thick gold film was sputtered onto a clean glass cover slip (thickness ≈ 150 μm). Slit and grooves structures were milled in the film by focused ion beam (FIB). One of the structure milled is shown in Fig. 1(b). A dielectric layer was then added above the slit and groove structures with the following procedure. 100 mg of Poly(methyl methacrylate) (PMMA) was dissolved in 1 mL anisole ($CH_3OC_6H_5$). The solution was spun onto the gold structure at a speed of 5300 revolutions per minute. This results in a 250 ± 10 nm thick PMMA layer above the metallic structure. This corresponds to the thickness suggested in Ref. 11. Both slit and grooves are completely filled with PMMA. The profile of the PMMA film on top of the structure was measured by atomic force microscopy (AFM). Fig. 1(c) shows this profile after averaging over several line scans taken perpendicular to the slit and groove direction. A modulation $x < 25$ nm is measured on top of the grooves, which can be considered as optically flat ($x < \lambda/10$). Above the slit, a groove of height $h = 100 \pm 10$ nm has been measured. This will have little influence on the optical properties of the structure as $\lambda/5 < h < \lambda/10$.

Transmission measurements were taken at normal incidence using a halogen light source. The low numerical aperture (NA < 0.1) incident beam is linearly polarized using a Glan laser prism (extinction ratio $10^5$). The transmitted light is collected with a microscope objective (NA = 0.6 and power 40×) and analyzed with a monochromator and a cooled charge-couple-device camera. The transmission through the structure is normalized to the photon flux incident on the open aperture (i.e. the slit).

In order to validate the experimental measurements, we have performed numeric calculations using the coupled mode method.[11] For simplicity, gold has been treated as a perfect electric conductor (PEC) and a fixed dielectric constant $\varepsilon_d$ = 2.25 has been used to represent both the PMMA layer and glass film. As a rule, the PEC approximation provides a good semi-quantitative agreement with the experiment for s-polarization in the optical region. For p-polarization the results based on PEC approach are not reliable for wavelengths shorter than 600 nm, but the agreement improves with increasing wavelength. The finite collection angle of the experiment has been taken into account in the calculations by restricting the integration interval of the propagating part of the transmitted angular spectra.

Transmission measurements were first made without the PMMA layer. For p-polarization, high transmission induced by SPP excitation is expected at a wavelength $\lambda$ close to $\lambda_{SPP}^{(n)} = (p/n)[\varepsilon_d \varepsilon_m / (\varepsilon_d + \varepsilon_m)]^{1/2}$ where $n$ is an integer indicating the SPP order, $\varepsilon_m$ and $\varepsilon_d$ are respectively the dielectric constant of the metal and the dielectric medium at the interface with the metal (in the present case air).[1,2] As $\lambda_{SPP}^{(n)}$ scales with $p$, the transmission peak is red-shifted on increasing $p$. This is what is seen in Fig. 2(a) for $w$ = 311 nm, where the high transmission peak ($\eta \approx 1$) measured at $\lambda \approx 725$ nm for $p$ = 600 nm shifts to $\lambda \approx 800$ nm for $p$ = 680 nm. In the s-polarization case, the transmitted spectrum is independent of period [Fig. 2(b)] since no surface waves are resonantly excited. At wavelengths larger than the cut-off wavelength $\lambda_c$, the fundamental slit mode for s-polarization is evanescent, inducing a transmission reduction[12,13]. $\lambda_c = 2w\varepsilon_d^{1/2}$ for perfect metals. $\lambda_c$ is slightly larger in the case of real metals due to their finite conductivity[13]. For $w$ = 311nm, the transmission drops for $\lambda$ > 700nm. Decreasing $w$ reduces the transmission[12], as is shown in Fig. 2(c). Measurements are in good agreement with the calculated spectra.

Transmission spectra for s-polarization with a dielectric layer added on top of the metal are shown in Fig. 3. The dielectric layer modifies the transmission properties of the structure, as expected from Ref. 11. Spectra are now dependent on periodicity. This indicates that the metallic grating allows incident light to resonantly couple into the dielectric layer. This coupling occurs at $\lambda_w^{(n)} = (p/n)q_w$, where $q_w$ is the effective index of the waveguide mode. Note that in the 250 nm dielectric layer, only one s-polarized waveguide mode is supported in the wavelength range considered.[11] High transmission peaks are measured, showing that the dielectric layer is acting as a waveguide and efficiently couples light through the slit.

The highest transmission measured is $\eta = 2.5$ for $p = 680$ nm. At $\lambda \approx 800$ nm, this corresponds to a 25 times increase as compared to the s-polarization case without PMMA layer. Note that the total transmission is higher than reported here as not all the light is collected due to the finite numerical aperture of the objective. Reducing the slit width such that $\lambda > \lambda_c$ drastically reduces light transmission [see Fig. 3(c)]. Even if the incident light is resonantly coupled into the waveguide mode, light is not transmitted by the single slit for $w = 200$ nm as the slit mode is evanescent and thus does not allow high transmission for a 420 nm thick film. In the p-polarization case, $\lambda_{SPP}^{(n)}$ is red-shifted due to the presence of the PMMA layer, thus displacing the peaks locations (not shown here due to space limitations).

Theory predicts twice the measured transmission [see Fig. 3(b) and 3(d)]. This quantitative discrepancy may be due to several reasons such as sample imperfections (e.g. PMMA layer flatness, gold roughness and structure profile) and PEC approximation in the calculations.

On the other hand, peak positions for the s-polarization case are predicted very accurately theoretically. This is explained by the fact that $\lambda_w^{(n)}$, which governs peak positions, is weekly affected by sample imperfections and approximations made in calculation. Note that for s-polarized waveguide modes, the field is not as tightly bounded to the surface as in case of SPP. Therefore, approximating a real metal to a perfect one predicts more accurately peak position for s than for p-polarization.

In conclusion, extraordinary optical transmission has been experimentally demonstrated for s-polarization. Adding on top of a slit and groove structure a thin dielectric film, which supports s-polarized waveguide modes, efficiently boosts s-polarized transmission. This experiment illustrates the fact that different kinds of surface waves can be used to enhance transmission. Good agreement is observed between experiment and the coupled mode method used to calculate the transmitted spectra.

It should be possible to transmit both polarizations at the same wavelength, thus increasing the overall transmission and removing polarization sensitivity. This would be particularly useful in optical devices such as low noise[8] or high speed photodetectors.[10]

This work was funded by the European Community, project no. IST-FP6- 034506 'PLEAS'. AYN acknowledges MICINN for a Juan de la Cierva Grant.

[1] D. E. Grupp, H. J. Lezec, T. Thio, and T. W. Ebbesen, Adv. Mater. **11,** 860 (1999).

[2] H. J. Lezec, A. Degiron, E. Devaux, R. A. Linke, L. Martín-Moreno, F. J. García-Vidal, and T. W. Ebbesen, Science **297,** 820 (2002).

[3] F. J. García-Vidal, H. J. Lezec, T. W. Ebbesen, and L. Martín-Moreno, Phys. Rev. Lett. **90,** 213901, (2003).

[4] O. T. A. Janssen, H. P. Urbach, and G. W. 't Hooft, Phys. Rev. Lett. **99,** 43902 (2007).

[5] H. Raether, *Surface Plasmons*, (Springer-Verlag, Berlin, 1988).

[6] J. B. Pendry, L. Martín-Moreno, and F. J. García-Vidal, Science **305**, 847 (2004).

[7] Y. Takakura, Phys. Rev. Lett. **86,** 5601 (2001).

[8] L. A. Dunbar, M. Guillaumée, F. Leon-Perez, C. Santschi, E. Grenet, R. Eckert, F. López-Tejeira, F. J. García-Vidal, L. Martín-Moreno, and R. P. Stanley, Appl. Phys. Lett. **95,** 011113 (2009).

[9] N. Yu, Q. J. Wang, C. Pflügl, L. Diehl, F. Capasso, T. Edamura, S. Furuta, M. Yamanishi, and H. Kan, Appl. Phys. Lett. **94,** 151101 (2009).

[10] T. Ishi, J. Fujikata, K. Makita, T. Babat, and K. Ohashi, Jpn. J. Appl. Phys. **44,** 364 (2005).

[11] A. Yu. Nikitin, F. J. García-Vidal, and L. Martín-Moreno, J. Opt. A: Pure Appl. Opt. **11,** 125702 (2009).

[12] M. Guillaumée, L. A. Dunbar, C. Santschi, E. Grenet, R. Eckert, O. J. F. Martin, and R. P. Stanley, Appl. Phys. Lett. **94,** 193503 (2009).

[13] H. F. Schouten, T. D. Visser, D. Lenstra, and H. Blok, Phys. Rev. E **67,** 36608 (2003).

FIG. 1. (Color online) (a) A schematic of the studied structure and definitions used in this letter for the polarization state and the structure dimensions. (b) SEM image of a slit and groove structure. (c) Profile of the PMMA layer deposited above a slit and groove structure measured by AFM.

FIG. 2. (Color online) Normalized transmission spectra of slit and groove structures milled in a 420 ± 10 nm thick gold film. Each slit is flanked by 7 periodic grooves 180 nm deep and 325 nm wide. Both slits and grooves are 10 µm long. Measurements for (a) p-polarization and $w$ = 311 nm, (b) s-polarization and $w$ = 311 nm, (c) s-polarization and $p$ = 638 nm. (d) Theoretical calculations from the coupled mode method for the same parameters than (c).

FIG. 3. (Color online) Normalized transmission spectra of slit and groove structures covered by a thin dielectric layer. The slit and groove dimensions are the same than for Fig. 2. [(a), (c)] s-polarized normalized transmission measurements; [(b), (d)] corresponding theoretical calculations from the coupled mode method. In [(a), (b)], $w$ = 311 nm; in [(c), (d)] $p$ = 638 nm.

# Figure 1

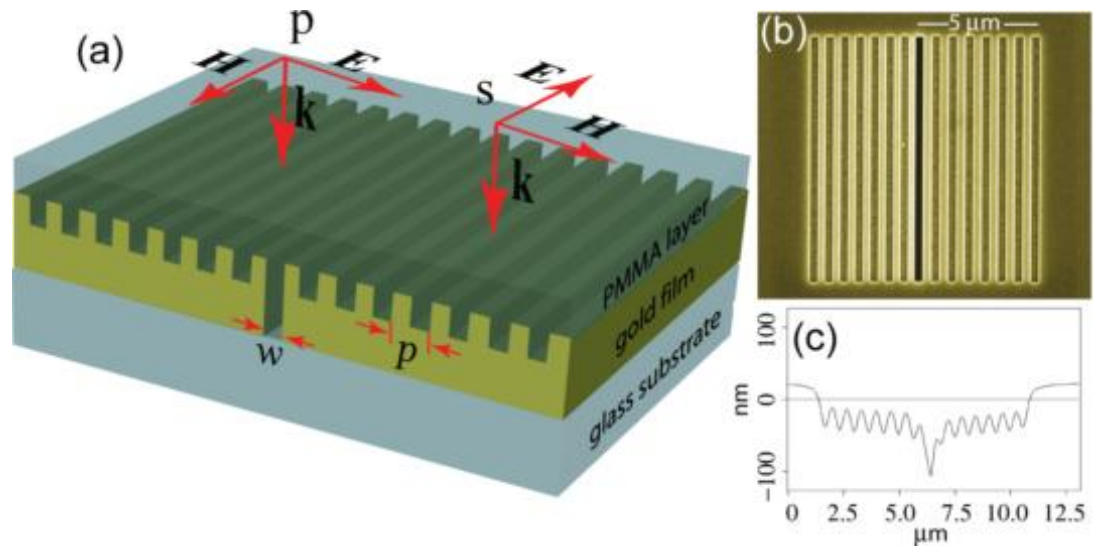

# Figure 2

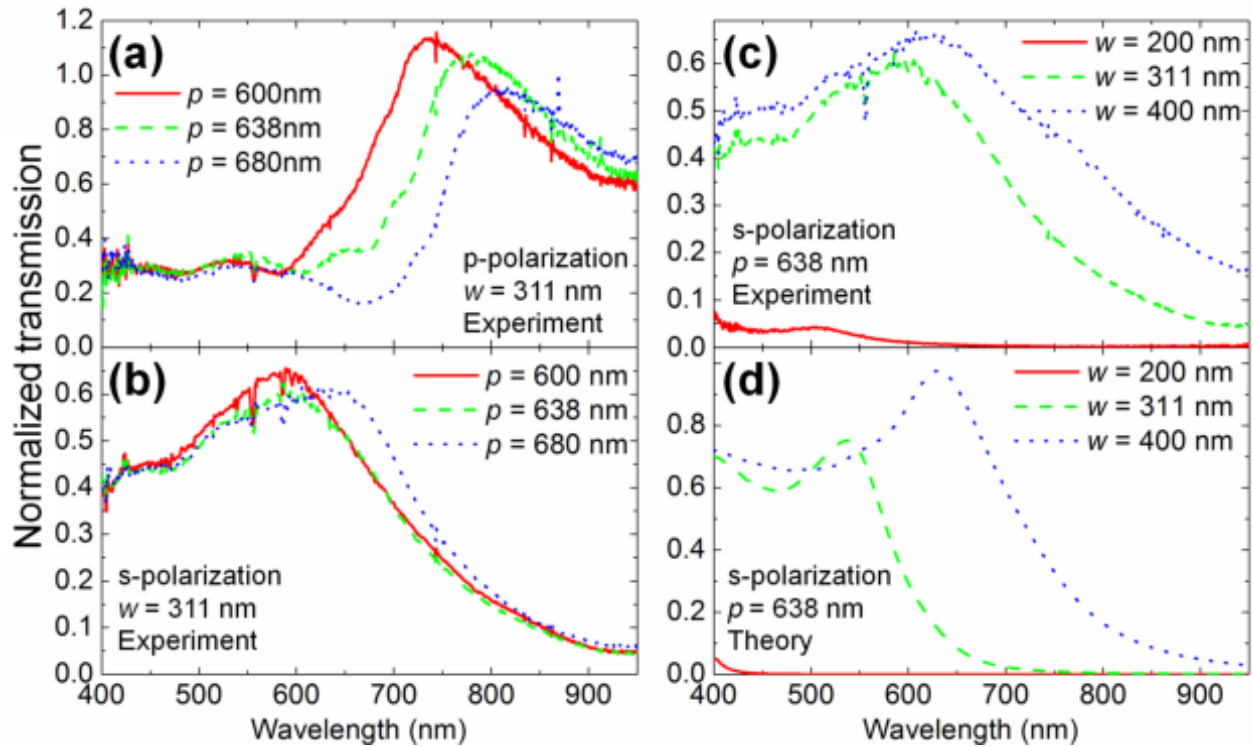

# Figure 3

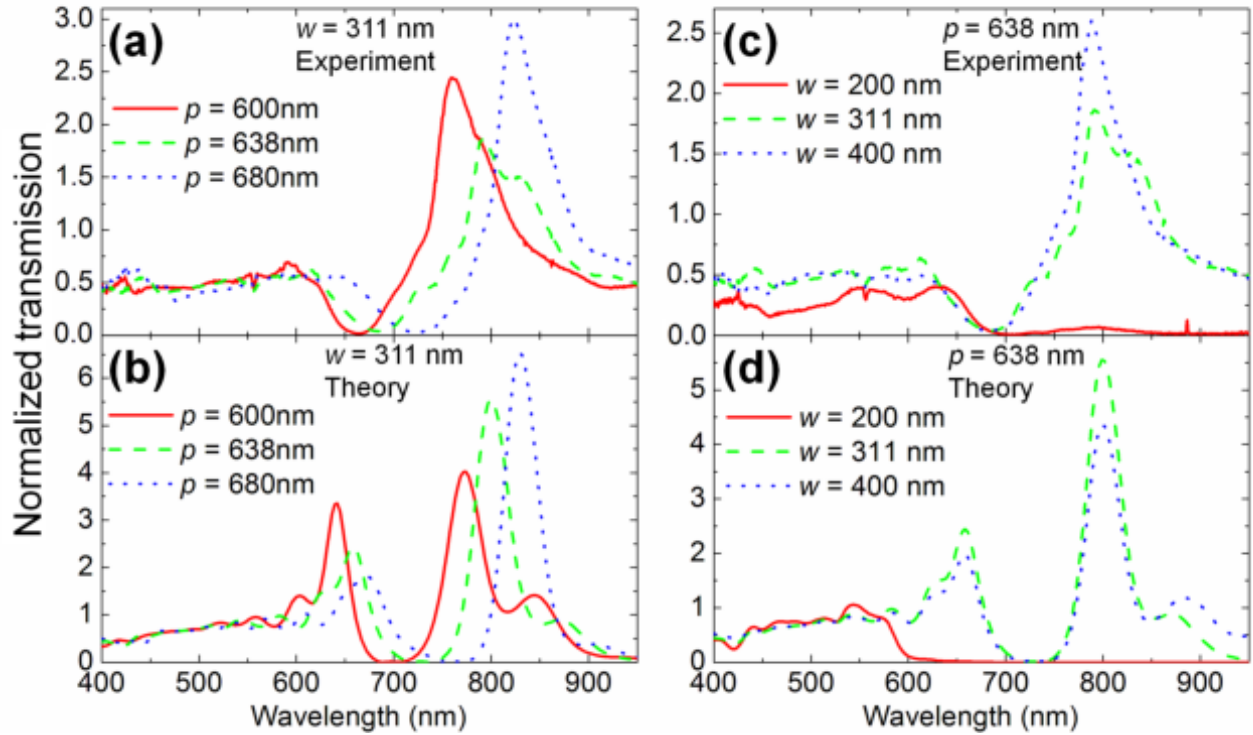